\documentclass[prl,twocolumn,aps,showpacs]{revtex4}

\usepackage[dvips]{graphicx}

\begin{document}

\title{2D Born-Infeld electrostatic fields}

\author{Rafael Ferraro}\thanks{Member
of Carrera del Investigador Cient\'{\i}fico (CONI\-CET,
Argentina).\\ Electronic address: ferraro@iafe.uba.ar\\ }

\affiliation{Instituto de Astronom\'{\i}a y F\'{\i}sica del
Espacio,Casilla de Correo 67, Sucursal 28, 1428 Buenos Aires,
Argentina,\\ and Departamento de F\'{\i}sica, Facultad de Ciencias
Exactas y Naturales, Universidad de Buenos Aires, Ciudad
Universitaria, Pabell\'on I, 1428 Buenos Aires, Argentina}

\date{September 18, 2003}

\begin{abstract}
The electrostatic configurations of the Born-Infeld field in the
2-dimensional Euclidean plane are obtained by means of a
non-analytical complex mapping which captures the structure of
equipotential and field lines. The electrostatic field reaches the
Born-Infeld limit value when the field lines become tangent to an
epicycloid around the origin. The total energy by unit of length
remains finite.
\end{abstract}


\maketitle

In 1934 Born and Infeld \cite{Born} proposed a non-linear
electromagnetism that modifies the behavior of the Maxwell theory
in the regime of strong fields. The aim was to formulate a theory
where the self-energy of a point-like charge is finite, and thus
open the possibility of conceiving a charged particle as a part of
the field instead of an external source of it. The basic idea was
to impose a finite limit value $b$ to a purely electrostatic
field. This could be achieved by reproducing the way the velocity
of a particle remains lower than $c$ when the classical Lagrangian
$L=(1/2)\,m\,\dot q^2$ is replaced by the relativistic Lagrangian
$L=-mc^2\,\sqrt{1-\dot q^2\,c^{-2}}$. This means that the Maxwell
Lagrangian $L_{M}=- \sqrt{-g}\, (8\pi c)^{-1}(B^2-E^2)$ should be
replaced by the Lagrangian $L_{B}=-\sqrt{-g}\, (4\pi
c)^{-1}b^{2}\,\sqrt{1+(B^{2}-E^{2})\,b^{-2}}$ \cite{B}. In order
that the energy goes to zero when the field goes to zero, a
``rest'' energy $\sqrt{-g}\,(4\pi c)^{-1}b^{2}$ should be
subtracted from $L_B$, without affecting the dynamical equations.
Nevertheless, Born and Infeld followed Einstein by judging that
the Lagrangian should combine the metric $g_{ij}$ and the
electromagnetic field $ F_{ij}\ =\ \partial _i A_j-\partial _j
A_i$ as the symmetric and antisymmetric parts of a unique field
$b\, g_{ij}+F_{ij}$. The Born-Infeld Lagrangian density is
\begin{eqnarray}
L_{BI}[A_{k}]\ &=-\frac{1}{4\pi c}\ \sqrt{\,|\det (b\
g_{ij}+F_{ij})|}\ + \sqrt{-g}\,\frac{b^{2}}{4\pi c}\cr\cr  &=\
\sqrt{-g}\ \frac{b^{2}}{4\pi c}\ \Big( 1- \sqrt{\
1+\frac{2S}{b^{2}}-\frac{P^{2}}{b^{4}}}\Big) , \label{Lagrangiano}
\end{eqnarray}
where $S$ and $P$ are the scalar and pseudoscalar field
invariants,
\begin{eqnarray}
S &=&  \frac{1}{4}\ F_{ij}F^{ij} =
\frac{1}{2}\,(B^{2}-E^{2})\,,\cr\cr P &=&  \frac{1}{8}\
\sqrt{-g}\,\varepsilon _{ijkl}\ F^{kl}\,F^{ij} = \frac{1}{4}\
^*\!F_{ij}F^{ij} = {\mathbf E}\cdot \mathbf{B}
\end{eqnarray}
$\varepsilon _{ijkl}$ being the Levi-Civita symbol whose
components are $\pm 1$ depending on $(ijkl)$ is an even or odd
permutation of $(0123)$. The Maxwell Lagrangian is recovered from
the Born-Infeld Lagrangian when $b\rightarrow \infty $. The field
equations derived from the Born-Infeld Lagrangian
(\ref{Lagrangiano}) are
\begin{equation}
\partial _{j}\ \left( \ \sqrt{-g}\ {\mathcal F}^{ij}\right) \, =\, 0
\label{fieldequations}
\end{equation}
where ${\mathcal F}_{ij}$ is the tensor
\begin{equation}
{\mathcal F}_{ij}\, = \, \frac{\ F_{ij}-b^{-2}\ P\,
^*\!F_{ij}}{\sqrt{\,1+\frac{2S}{b^{2}}-\frac{ P^{2}}{b^{4}}}}
\end{equation}
Since the field is an exact 2-form ($F=dA$), the identities $dF=0$
must be added to the Euler-Lagrange equations
(\ref{fieldequations}).

The energy-momentum tensor results (the metric signature is
$(+---)$)
\begin{eqnarray}
T_{ij}=\frac{2c}{\sqrt{-g}}\, \frac{\partial L_{BI}}{\partial g^{i
j}}&\cr\cr =- \frac{1}{4\pi}\,F_{ik}\,\,{\mathcal F}_j^{\,\,\,k}\
-&\frac{b^2}{4\pi} \ g_{ij}\left(
1-\sqrt{1+\frac{2S}{b^2}-\frac{P^2}{b^4}}\right)
\label{energia-momento}
\end{eqnarray}

To get the Born-Infeld charge we solve Eq.(\ref{fieldequations})
for an isotropic electrostatic field $F=E(r)\,dt\wedge dr$. Since
$\sqrt{-g} =r^2\sin \theta $ in spherical coordinates, the
solution is
\begin{eqnarray}
{\mathcal F}=& q\ r^{-2}\,\,dt\wedge dr\,,\,\,\,\,\,F=\,b\
\Big(\frac{b^2\,r^4}{q^2}+1\Big)^{-\frac{1}{2}}\,\,dt\wedge
dr\,,\cr\cr &T_{0}^{\,\,\,0}=\frac{b^{2}}{4\pi
}\sqrt{1+\frac{q^{2} }{b^{2}r^{4}}}-\frac{b^{2}}{4\pi }\,,\cr\cr
&U=\int_{0}^{\infty }T^{00}\,4\pi
r^{2}dr\,=\frac{1}{6}\sqrt{\frac{q^{3}b}{\pi }}\,\,\Gamma (
\frac{1}{4})^{2} \label{electric}
\end{eqnarray}

While $F$ and ${\mathcal F}$ are equal in Maxwell theory, they
differ in the Born-Infeld theory. As a result, only one of them
diverges: ${\mathcal F}$ diverges at $r_{o}=0$, but
$F(r_{o})=b\,\,dt\wedge dr$. This characteristic moderates the
divergence of the energy-momentum tensor, and leads to a finite
energy $U$.

Another nice example is the axial magnetostatic field $F=B(\rho
)\,dz\wedge d\rho $. Since $\sqrt{-g}=\rho $ in cylindrical
coordinates, then the solution of Eq.(\ref{fieldequations}) is
\begin{eqnarray}
{\mathcal F}=\,\frac{2I}{c\rho }\,\,dz&\wedge d\rho\,,
\;\;\;\;\;F=\, b\ \Big(\frac{b^2\,\rho ^2
c^2}{4\,I^2}-1\Big)^{-\frac{1}{2}}\,\,dz\wedge d\rho\,,\cr\cr
&T_{0}^{\,\,0}=\frac{b^2}{4\pi}\,\Big( 1-\frac{4\,I^2}{b^2\,\rho^2
c^2}\Big)^{-\frac{1}{2}}-\frac{b^{2}}{4\pi }\label{magnetic}
\end{eqnarray}

Now $F$ diverges at $\rho _{o}=\frac{2\,I}{c\,b}$, but ${\mathcal
F}(\rho _{o})=b\,\,dz\wedge d\rho $. Although the energy is not
finite in this case, as a consequence of the extended character of
the source, however the integral $\int T^{00}\,2\pi \rho \,d\rho $
remains finite at $\rho =\rho _{o\,}$.

In the last decades there was a renewal of interest in the
Born-Infeld theory because it emerges in the low energy limit of
string theories \cite{Frad,Berges,Met,Tsey,Gaunt,Brech}. Maxwell
and Born-Infeld theories have proved to be the sole theories for
the massless spin 1 field having causal propagation
\cite{Pleb,Deser} and absence of birefringence
\cite{Boillat,Bialy}. However the essential features of field
configurations other than the kind above considered are hardly
known, due to the problem of dealing with the non-linear equations
involved in the theory. Here we are going to introduce a procedure
that works for Born-Infeld electrostatic fields lying in the
2-dimensional plane. This procedure extends the method of using
analytic complex functions to get solutions of the Laplace
equation in 2 dimensions. As it is well known, if $w({\mathbf
z})=u(x,y)+i\,v(x,y)$ is an analytical function in the complex
plane, then $u(x,y)$ and $v(x,y)$ solve the Laplace equation.
Analytic functions generate conformal mappings $ {\mathbf z}=f(w)$
in the Euclidean plane; in fact, $dx^{2\,}+dy^{2}=dz\
dz^{*}=f^{\prime }\,f^{\prime *}dw\, dw^{*}=\left| f^{\prime
}(w)\right| ^{2}(du^{2\,}+dv^{2})$, so $u,v$ are orthogonal
coordinates dilating distances without changing the shapes of
infinitesimal figures. If we regard $u(x,y)$ as the electrostatic
potential, then the coordinate lines $ u(x,y)=const.$ are
equipotential and the coordinate lines $v(x,y)=const.$ are field
lines.

Although the Born-Infeld electrostatic potential is not a solution
of the Laplace equation, a modified version of the complex mapping
can be still applied to get the structure of the Born-Infeld
electrostatic field in 2 dimensions. The substitute mapping must
generate orthogonal coordinates $u,v$ --the field lines are
orthogonal to the equipotential surfaces--, but it will distort
the infinitesimal shapes. So let us try with
\begin{equation}
{\mathbf z}=f(w)+\frac{g(w^{*})}{4\ b^{2}} \label{mapping}
\end{equation}
where $f$ and $g$ are analytic functions of their respective
arguments (in the sense that $df/dw^{*}=0$ and $dg/dw=0$).
Besides, let us choose
\begin{equation}
f^{\prime }(w)g^{\prime }(w^{*})^{*}=1 \label{mapping2}
\end{equation}
Then
\begin{eqnarray}
d{\mathbf z}&=&f^{\prime }(w)\ dw + \frac{g^{\prime
}(w^{*})}{4b^{2}}\ dw^{*}\cr\cr &=&f^{\prime }(w)\
dw+\frac{1}{4b^{2}f^{\prime }(w)^{*}}\ dw^{*} \cr\cr &=&f^{\prime
}(w)\Big[\Big( 1+\frac{\left| f^{\prime }(w)\right|
^{-2}}{4b^{2}}\Big)\ du \cr\cr &&+\ i\ \Big(1-\frac{\left|
f^{\prime }(w)\right| ^{-2}}{4b^{2}} \Big)\
dv\Big]\label{coordinate change}
\end{eqnarray}
and
\begin{eqnarray}
&dx^2&+dy^2=d{\mathbf z}\ d{\mathbf z}^{*}=\left| f^{\prime
}(w)\right|^{2}\Big[ \Big( 1+\frac{\left| f^{\prime
}(w)\right|^{-2}}{4b^{2}}\Big)^{2} du^{2}\cr\cr &+& \Big(
1-\frac{\left| f^{\prime }(w)\right|^{-2}}{4b^{2}}\Big)
^{2}dv^{2}\Big] =-g_{uu}\,du^{2}-g_{vv}\,dv^{2}
\end{eqnarray}

\begin{figure*}[t]
\includegraphics[width=15cm]{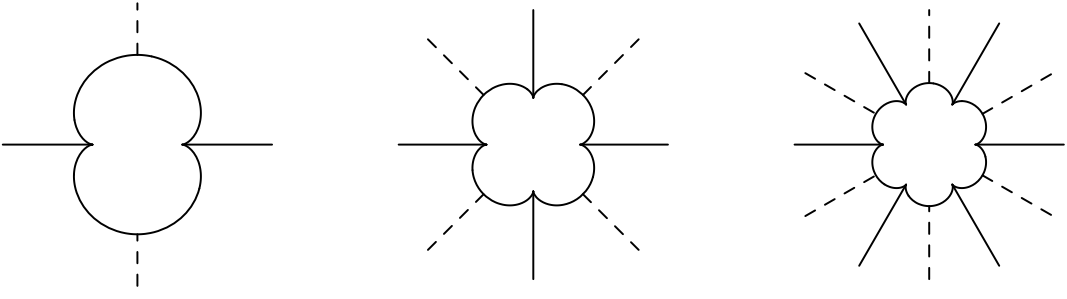}
\caption{Epicycloids with 2, 4 and 6 cusps. Lines $u\,=\,0$
(dashed lines) and $v\,=\,0$.} \label{epicycloids}
\end{figure*}

So the mapping (\ref{mapping})-(\ref{mapping2}) effectively
generates orthogonal coordinates $u,v$ in the Euclidean plane. Now
we will prove that $u(x,y)$ is the electrostatic potential of a
Born-Infeld field, i.e. the exact 2-form $F=du\wedge dt$ solves
the Born-Infeld equations. Since $2S$ results to be $g^{uu}$, then
\begin{equation}
\sqrt{-g}\ {\mathcal F}^{\,t\,u}\
=-\frac{\sqrt{g_{uu}g_{vv}}\,\,g^{uu}}{
\sqrt{1+\frac{g^{uu}}{b^{2}}}}=\frac{\sqrt{g^{uu}g_{vv}}\,\,}{\sqrt{1+\frac{
g^{uu}}{b^{2}}}}=1
\end{equation}
for all $f^{\prime }(w)$, and the Eq.(\ref{fieldequations}) is
fulfilled. The cartesian components of the electric field,
$E_{x}=-\partial u/\partial x,\,E_{y}=-\partial u/\partial y$, are
obtained by inverting the Jacobian matrix of the coordinate
transformation (\ref{mapping})-(\ref{mapping2}),
\begin{eqnarray}
\frac{\partial x}{\partial u}&=&\Big( 1+\frac{\left| f^{\prime
}(w)\right| ^{-2}}{ 4b^{2}}\Big)\ {\mathrm Re} f^{\prime }(w)
\cr\cr \frac{\partial y}{
\partial u}&=&\Big( 1+\frac{\left| f^{\prime
}(w)\right| ^{-2}}{4b^{2}}\Big)\ {\mathrm Im} f^{\prime }(w)
\cr\cr \frac{\partial x}{\partial v}&=&-\Big( 1-\frac{\left|
f^{\prime }(w)\right| ^{-2}}{ 4b^{2}}\Big)\ {\mathrm Im} f^{\prime
}(w) \cr\cr \frac{\partial y}{\partial v }&=&\Big( 1-\frac{\left|
f^{\prime }(w)\right| ^{-2}}{4b^{2}}\Big)\ {\mathrm Re} f^{\prime
}(w)
\end{eqnarray}

The inverse matrix is
\begin{eqnarray}
-E_{x} =\frac{\partial u}{\partial x}&=&\frac{4b^{2}\ {\mathrm
Re}f^{\prime }(w)}{ 1+4b^{2}\left| f^{\prime }(w)\right| ^{2}}
\cr\cr \frac{\partial v}{\partial x}&=&\frac{4b^{2}\ {\mathrm Im}
f^{\prime }(w)}{1-4b^{2}\left| f^{\prime }(w)\right| ^{2}} \cr\cr
-E_{y} =\frac{\partial u}{\partial y}&=&\frac{4b^{2}\ {\mathrm
Im}f^{\prime }(w)}{ 1+4b^{2}\left| f^{\prime }(w)\right| ^{2}}
\cr\cr \frac{\partial v}{\partial y}&=&-\frac{4b^{2}\ {\mathrm Re
}f^{\prime }(w)}{1-4b^{2}\left| f^{\prime }(w)\right| ^{2}}
\end{eqnarray}

and the field is
\begin{equation}
{\mathbf E}=E_{x}+iE_{y}=-\frac{4b^{2}f^{\prime
}(w)}{1+4b^{2}\left| f^{\prime }(w)\right| ^{2}}\label{field}
\end{equation}

In spite of the simple appearance of this result for electrostatic
fields in 2D, it must be remarked that the difficulty lies in
writing $w$ at $ f^{\prime }(w)$ as a function of $(x,\,y)$,
because this implies solving the Eq.(\ref{mapping}).

Special interest deserve the periodic non-isotropic solutions. In
Maxwell theory these are $u_{M}(x,y)$ $=A\,\rho ^{-n}\cos n\varphi
,\,n\in {\mathrm N} $, and come from the mapping ${\mathbf
z}=f(w)=A\,w^{-\frac{1}{n}}$, where $A$ is a constant.
Consequently, for the Born-Infeld field we will use the mapping
\begin{equation}
{\mathbf z}=\frac{A}{w^{\frac{1}{n}}}-\frac{w^{*\
\frac{1}{n}+2}}{4\frac{1}{n} (\frac{1}{n}+2)\ A\,b^{2}}
\end{equation}

In order to check the periodicity of this mapping, let us note
that the lines $u$ $=0$ (i.e. $w=i\,v$) are the points where
$i^{1/n}\,{\mathbf z}$ is real. Thus the lines $u$ $=0$ coincide
with those of the Maxwell field, so guaranteeing the periodicity
of the mapping (the same can be said about the lines $v$ $=0$).
Our first task is to find the places where the field reaches the
limit value $b$. According to Eq.(\ref{field}), this happens where
$ |f^{\prime }(w)|=(2\,b)^{-1}$. Therefore
\begin{equation}
\frac{1}{2\,b}=\left| f^{\prime }(w)\right| =\frac{A}{n}\,\left|
w\right| ^{- \frac{1}{n}-1}
\end{equation}
This equation corresponds to a circle in coordinates $u,v$, which
can be parameterized as $w(\tau
)=(\frac{2}{n}\,A\,b)^{\frac{n}{n+1}}\,\exp (-in\tau )$. By
replacing this parametrized curve in the expression for ${\mathbf
z}$ , we get parametric equations in cartesian coordinates:
\begin{eqnarray}
{\mathbf z}(\tau )=x(\tau )+i\,y&(\tau)&
\,=\,A\,\Big(\frac{2}{n}\,A\,b\Big)^{-\frac{1}{ n+1}}\,\Big( \exp
(i\,\tau )\cr\cr  &-&\frac{1}{1+2n}\,\exp [i\,(1+2n)\tau ]\Big)
\end{eqnarray}

This curve is a $2n$-cusped epicycloid, which is represented in
Fig.~\ref{epicycloids} together with the lines $u$ $=0,\,v=0$. The
cusps are the points where $\tau =k\pi/n,\,k\in {\mathrm Z}$; then
${\mathbf z}_{k}=$ $\frac{2n}{1+2n}
\,A\,\,(\frac{2}{n}\,A\,b)^{-\frac{1}{n+1}}\,\exp (i\,k\pi/n)$.
$w=u+i\,v$ is real at the cusps; so $|u|$ reaches there its
maximum value $(\frac{2 }{n}\,A\,b)^{\frac{n}{n+1}}$. The cusps
should not be regarded as point-like charges because the field
lines do not converge on the cusps. Instead, the field lines
tangentially reach the epicycloid. In fact, on the one hand the
complex field ${\mathbf E}$ on the epicycloid is ${\mathbf E}(\tau
)=-2\,b^{2}\,f^{\prime }(w(\tau ))=-2\,b^{2}\,\frac{A}{n}\,w(\tau
)^{-\frac{1 }{n}-1}=-b\,\exp [i\,(1+n)\tau ]$. On the other hand
the vector tangent to the epicycloid is $d{\mathbf z}/d\tau
=i\,A\,\,(\frac{2}{n}\,A\,b)^{-\frac{1}{ n+1}}\,\left( \exp
(i\,\tau )-\exp [i\,(1+2n)\tau ]\right) =2\,A\,\,(\frac{2
}{n}\,A\,b)^{-\frac{1}{n+1}}\,\sin n\tau \,\exp [i\,(1+n)\tau ]$,
which is parallel to ${\mathbf E}(\tau )$.

\begin{figure*}[t]
\includegraphics[width=15cm]{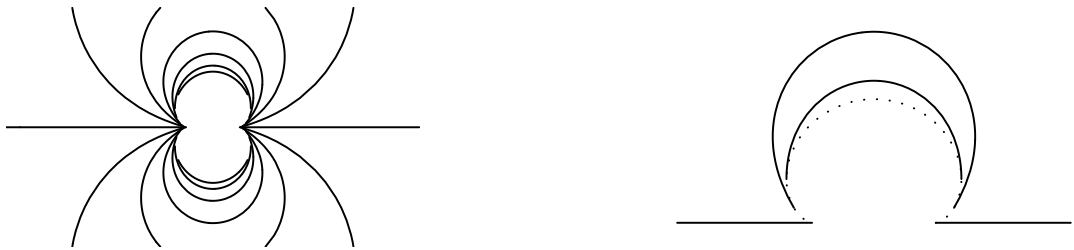}
\caption{Born-Infeld field lines for $f(w)\,\propto w^{-1}$.
Behavior of the field near the nephroid.}\label{n=1field}
\end{figure*}

In Fig.~\ref{n=1field} we show the main features of the 2D
Born-Infeld electrostatic field for the case $n=1$. In Maxwell
context, this case corresponds to a pair of infinitely close
parallel opposite uniform line charges, and $A$ is the separation
distance times the linear density of charge. The 2-cusped
epicycloid is a nephroid. The mapping is
\begin{eqnarray}
x\,&=&\,{\mathrm Re}\big[Aw^*\big(\frac{1}{w w^*}-\frac{w^{*\,
2}}{12\,A^2 b^2}\big)\big]\cr\cr  &=&\frac{A\,u}{u^2+v^2}-u\
\frac{u^2-3\,v^2}{12\, A b^{2}}\cr\cr  y\,&=&\,{\mathrm
Im}\big[Aw^*\big(\frac{1}{w w^*}-\frac{w^{*\, 2}}{12\,A^2
b^2}\big)\big]\cr\cr  &=&-\frac{A\,v}{u^2+v^2}-v\
\frac{v^2-3\,u^2}{12\, A b^{2}}
\end{eqnarray}

Since the coordinate lines $v=v_{o}=const.$ are field lines, then
the equations $x=x(u,v_{o}),\,y=y(u,v_{o})$ are parametric
equations for the field lines and can be numerically plotted. The
equipotential lines are plotted in the same way.

The energy of the field is the integral of $T^{\,00}$ outside the
epicycloid. This integral gets its simplest form when it is
expressed in terms of coordinates $u,\,v$ because the integration
region is a circle of radius $(\frac{2}{n}\,A\,b)^{\frac{n}{
n+1}}$ (remember the epicycloid is a circle in coordinates $u,v$;
the field lives inside the circle because $ u,v$ goes to zero at
the infinity). Since the volumen is $\sqrt{g_{uu}g_{vv}}
=-g_{uu}\,\sqrt{\,1+\frac{g^{uu}}{b^{2}}}=E^{-2}\,\sqrt{\,1-\frac{E^{2}}{
b^{2}}}$, then
\begin{eqnarray}
U&=&\int T^{\,\,00}\,\sqrt{\,g_{uu}\,g_{vv}}\ du\,dv\,dz\cr\cr
&=&\frac{1}{4\pi }\,\int \,\frac{b^{2}}{E^{2}}\,\,\Big(
1-\sqrt{\,1-\frac{E^{2}}{b^{2}}}\ \Big)\ du\,dv\,dz\cr\cr
&=&\frac{1}{8\,\pi }\,\int \,\Big( 1+\frac{\left| f^{\prime
}(w)\right| ^{-2}}{4b^{2}}\,\Big)\ du\,dv\,dz\cr\cr
&=&\frac{1}{8\,\pi }\,\int \,\Big(
1+\frac{n^{2}\,(u^{2}+v^{2})^{1+\frac{1}{n}}}{4\,A^{2}b^{2}}
\,\Big)\ du\,dv\,dz
\end{eqnarray}
and the result is
\begin{equation}
U=\frac{1}{8}\,\frac{\,1+3n\,}{\,1+2n\,}\,\,\Big(\frac{2}{n}\,A\,
b\Big)^{\frac{2\,n}{\,n+1}} \int dz
\end{equation}

Although it is nice to find that 2D electrostatic Born-Infeld
fields have a finite energy by unit of length, some other features
of these fields seem to be less pleasant. The Euler-Lagrange
equations break down on the epicycloid because the tensor
${\mathcal F}$ diverges there, which prevents us integrating the
equations beyond the epicycloid; essentially the same thing
happens to the magnetostatic field of Eq.(\ref{magnetic}). The
field (\ref{electric}) of a point-like charge is not devoid of
problems since it is perplexing the finite value at the origin of
its isotropic vector field (Hoffmann and Infeld proposed a
modification of the Born-Infeld Lagrangian to avoid this behavior
\cite{HI}). Perhaps there is nothing wrong with these features,
but they only invite us to consider non trivial combinations of
electrostatic and magnetostatic fields as meaningful static
solutions. Since the theory is non linear, the static solutions
with both types of field do not reduce to a mere superposition. It
would be enjoyable that a point-like charge get a completely
satisfactory 3D monopolar electrostatic field once its Born-Infeld
field includes the dipolar magnetostatic component.

\bigskip

This work was partially supported by Universidad de Buenos Aires
(Proy. UBACYT X805)

\end{document}